\documentclass[epj,nopacs]{svjour}
\usepackage{graphicx}
\usepackage{xspace}

\usepackage[colorlinks=true,citecolor=blue,urlcolor=blue]{hyperref}
\usepackage{breakurl}

\newcommand{\Bbar}{\,\overline{\!B}{}}
\newcommand{\Dbar}{\,\overline{\!D}{}}
\newcommand{\Kbar}{\,\overline{\!K}{}}
\def\B0bar{\Bbar{}^0}
\def\D0bar{\Dbar{}^0}
\def\K0bar{\Kbar{}^0}

\def\FCCee{FCC-$ee$\xspace}
\def\fccee{\FCCee}

\makeatletter
\g@addto@macro\bfseries{\boldmath}
\makeatother

\begin{document}

\title{Theoretical challenges for flavor physics}

\author{Yuval Grossman\inst{1} \and Zoltan Ligeti\inst{2}}

\institute{
Department of Physics, LEPP, Cornell University, Ithaca,
NY 14853, USA, \email{yg73@cornell.edu} \and 
Lawrence Berkeley National Laboratory, University of California, Berkeley,
CA 94720, USA, \email{ligeti@lbl.gov} }

\date{}

\abstract{
We discuss some highlights of the \FCCee flavor physics program. 
It will help to explore various
aspects of flavor physics: to test precision calculations, 
to probe nonperturbative QCD methods,
and to increase the sensitivity to physics beyond the standard model. In some areas, \fccee will do much better than current and near-future experiments. We briefly
discuss several probes that can be
relevant for maximizing the gain from the \fccee flavor program.}

\maketitle

\section{Introduction}
\label{sec:intro}

The goal of the \FCCee program in its tera-$Z$ phase is to produce about $5\times 10^{12}$ $Z$ decays
(per experiment), which will greatly improve precision electroweak tests of the standard model (SM). 
These decays will yield about $10^{12}$ $b\bar b$ and $c\bar c$ pairs, as
well as a large and clean sample of $\tau^+\tau^-$ pairs.
Using these data, the \fccee can shed light on open issues in flavor
physics. The flavor physics capabilities of circular $e^+e^-$
colliders were recently reviewed in Refs.~\cite{Abada:2019lih,CEPCStudyGroup:2018ghi}.

There are several ways the \fccee can probe flavor physics: by directly 
producing heavy particles ($Z$, $W$, and $t$) and studying their properties, or by making highly sensitive measurements of decays of hadrons.
Using the production of the gauge bosons, we can directly probe
the flavor structure of their couplings.
For example, by measuring $Z$ decays we can test flavor
universality to very high precision both in the lepton and the quark
sectors.
Moreover, collecting $10^8$ $WW$ pairs would yield a qualitatively new
determination of $|V_{cb}|$ from $W\to b\bar c$ decays, with
\mbox{0.3\%\,--\,0.4\%}
uncertainty~\cite{MarieHelene,Azzurri}. Such a determination will be
independent of $|V_{cb}|$ measurements in $B$
decays.
(The statistical uncertainty of extracting $|V_{ub}|$ from $W\to  b
\bar u$ is estimated around $5\%$~\cite{Azzurri}, and would 
need to improve to be competitive with anticipated prior results.)
Also, collecting 1.5\,ab$^{-1}$ data near the $t\bar t$ threshold yields a
clean sample of $10^6$ $t\bar t$ events~\cite{Abada:2019lih}.  For some flavor-changing neutral-current (FCNC) top
decays, $t\to \{H, Z, \gamma\}\, q$, the sensitivity improves compared
to the HL-LHC.

In this brief essay we focus on other ways the \fccee can probe
the flavor sector of the SM.  We concentrate on the tera-$Z$ phase
of the \FCCee, and discuss its
capability to shed light on open issues in flavor physics,
especially on bottom and charm quark physics.
In particular, we aim to minimize the overlap with 
Refs.~\cite{Abada:2019lih,CEPCStudyGroup:2018ghi}, and focus on
less often discussed physics topics, to show that the \fccee
could be useful in many ways that are not yet fully developed.

In the next few years, the two main experiments in flavor
physics will be LHCb and Belle~II. (By LHCb we refer
throughout this article to all LHC experiments, as LHCb is expected to
dominate most flavor physics measurements, though in some modes ATLAS and CMS
will also contribute significantly.) Each of them can teach us
new things about flavor physics, that are partly overlapping and partly complementary. The \fccee program will take place later, and therefore it has
to be viewed in light of earlier findings. 

We do not know what the status of particle physics will be when the \fccee 
starts operating. 
If a deviation from the SM is established before the
\fccee starts, the physics program will be tuned to explore in detail that
direction of beyond standard model (BSM) physics. We assume in
this review  that no conclusive deviations from the SM will be
found before the \fccee.
Under that assumption, besides the fact that the \fccee will provide a
huge amount of data, the point is to exploit the 
unique capabilities of the \fccee compared to LHCb~\cite{Bediaga:2018lhg} and
Belle~II~\cite{Kou:2018nap}.

The \fccee and Belle~II share the clean environment provided by $e^+e^-$ colliders. Compared to Belle~II (with 50/ab), much more data is anticipated at the
\fccee, by roughly a factor of 10 (see Table~7.1 in 
Ref.~\cite{Abada:2019lih}).
Moreover, the \fccee also produces $b$-baryons and $B_s$ mesons (Belle~II
can also produce $B_s$ at the $\Upsilon(5S)$ resonance, but
much fewer than the \fccee, and cannot resolve $B_s$ oscillations.) Concerning $CP$ violation (CPV) in $B$ decays, at the \fccee the $b$ and $\bar b$ quarks hadronize independently (like at the LHC), while  Belle~II is an asymmetric collider in order to study the 
time-dependence of the decays of correlated $B$ mesons from $\Upsilon(4S)$ decays.

Compared to LHCb, there are important differences due to triggering. 
At the LHC, complex triggers make the data sets manageable, 
while the \fccee (and also Belle~II) can have a much more open trigger. 
For fully reconstructed decays to charged particles, the \FCCee
sensitivities will not be much better than at LHCb, so we focus on
channels which are hard for LHCb.  Since efficiencies at LHCb depend
hugely on the decay channels, one cannot make simple comparisons
based only on luminosity.
The clean environment of the \fccee will result in much
better sensitivities, for example, for final states involving neutrinos or neutral
mesons, that are not easily accessible at LHCb.
Moreover, since the initial state
is $CP$ symmetric at the \fccee, there are no production
asymmetries. This eliminates a systematic uncertainty
at the LHC experiments, which may become important
for some $CP$ asymmetry measurements as their 
sensitivities approach the per mille level.

Another important difference between the \fccee compared to both
Belle~II and the LHC is that quarks from $Z$ decays are highly polarized. 
In particular, the $b$ and $c$ quarks polarizations are very
large. These large values make 
studies that require polarization ideal for the \fccee.

\section{Specific probes}
\label{sec:bsm}

\subsection{$CP$ violation and BSM physics in $B_{d,s}$ meson mixing}
\label{sec:bsm.mix}

Among the three types of $CP$ violation, CPV in decay and CPV in the
interference of decay with and without mixing have been well
established in many $b$ hadron decays.  However, CPV in mixing has
only been observed in $K^0$ mesons so far.  Due to the many more decay
channels available in $B_{d,s}^0$ mesons,
CPV in $B$ mixing provides a complementary probe of BSM~\cite{Laplace:2002ik}.
It can be measured via the $CP$ asymmetry in semileptonic $B$
decays, $A_{\rm SL}$.

The current world averages for the $B_d$ and $B_s$ mesons are $A_{\rm SL}^d = -(2.1 \pm 1.7) \times 10^{-3}$ and 
$A_{\rm SL}^s = -(0.6 \pm 2.8) \times 10^{-3}$~\cite{Amhis:2019ckw}, respectively. These uncertainties are well above the SM expectations,
$A_{\rm SL}^d = -(4.7\pm0.6)\times 10^{-4}$ and $A_{\rm SL}^s =
(2.22\pm0.27)\times 10^{-5}$~\cite{Jubb:2016mvq}.  (The theory
uncertainties are subject to some recent discussions~\cite{Lenz:2020efu}.)
At the \fccee, the achievable
experimental uncertainty has been estimated to be about $2.5 \times 10^{-5}$ for both quantities~\cite{Monteil,Charles:2020dfl}, which would allow a measurement of $A_{\rm SL}^d$ even at the SM level. These measurements would help probe BSM physics, as well as help discriminate between models, should BSM physics be discovered in other processes.

In a large class of models, the dominant BSM physics effects in the flavor sector may be
those that modify neutral meson mixing amplitudes, while impacts on
decay rates may be smaller.  In addition to exploring specific models,
this is justified, or maybe even expected, from an effective field
theory viewpoint.  The scale of dimension-6 operators contributing to
neutral meson mixing is typically constrained to be higher by the data
than the scales of operators affecting decays. A recent analysis of future sensitivities observed that beyond the Belle~II and LHCb data taking in this decade, future improvements are hindered by the expected uncertainty of $|V_{cb}|$~\cite{Charles:2020dfl}.  To go beyond current expectations and reach the per mille level, not yet known theoretical progress would be needed.

\subsection{$CP$ violation in hadronic $b$ decays}

Measurements of $CP$ asymmetries in $b$ decays shaped our understanding that
the breaking of $CP$ symmetry observed in hadron decays is due to a single complex parameter of the CKM
matrix. Currently, there is a lot of effort at the LHC and Belle~II to
test the CKM picture of CPV to much higher precision. The drive for
this program is the fact that some observables are theoretically extremely
clean, and thus any experimental progress will improve the sensitivity to BSM physics. At the
same time, measurements of observables which can currently only be computed with large theoretical
uncertainties due to hadronic physics, provide us with
information about QCD. 

The \fccee is expected to deliver a very large and clean sample of
$b$ hadrons. It is anticipated that the uncertainties of the CKM angle 
$\gamma$ will reach about 0.004\,rad, of $\beta$ about 0.005, and of $\phi_s$ about 0.002 (see Table~7.3 of Ref.~\cite{Abada:2019lih}).
While these are only modest improvements over what is expected from LHCb (with 300/fb), 
these observables relate to measurements which can be done well at the LHC.
To interpret measurements of $\beta$ and $\phi_s$ with such precisions, improvements in the theory are also needed to relate the results to parameters in the Lagrangian.
The \fccee should have greater advantages in modes containing neutrals, but few dedicated sensitivity studies exist so far. 
It will also allow combining results from experiments in different environments, and thus with different systematic uncertainties. 

While most probes of CPV use rate asymmetries (with or without
time-dependence) one other way to probe $CP$ violation is to use angular distribution
asymmetries. They are usually called ``triple products'', which is an
idea that can be applied to many different angular
distributions~\cite{Durieux:2015zwa}.
The \fccee can be used to measure many such modes. A theoretical
question is how clean information can be extracted from them. The
result is sensitive to hadronic matrix elements in a way similar to
rate asymmetries. While it is interesting to measure them as a way to
get insights into QCD, it would be even more significant to find a way to
cleanly probe also the underlying weak interactions in such
cases. The hope is that some progress in this direction will take place
by the time the \fccee starts collecting data.

\subsection{Very rare decays}
\label{sec:bsm.rare}

The \FCCee will have unique capabilities to measure decay modes with large
missing energy.  These include decays with neutrinos (or $\tau$ leptons) in the final state.  The \fccee should also be able to measure electrons better than LHCb.  Both exclusive and inclusive measurements are interesting, as the experimental and theoretical uncertainties are distinct, so they provide complementary probes of the underlying physics.  

Prime examples are decays mediated by $b\to s\nu\bar\nu$ or $b\to
s\tau^+\tau^-$ transitions, as well as their $b\to d$ counterparts.  
While LHCb is well suited to measure 
$B\to K^{(*0)}\mu^+\mu^-$ with high precision, its $b\to d$ analogs, 
$B\to \rho\mu^+\mu^-$ or $B_s\to K^{(*0)}\mu^+\mu^-$ are much more challenging~\cite{Aaij:2018jhg,MHS}.
The \fccee is expected to be able to tackle these, as well as
$B\to K^{(*0)}\tau^+\tau^-$, $\Lambda_b \to \Lambda\tau^+\tau^-$~\cite{Abada:2019lih},
$B\to K^{(*)}\nu\bar\nu$, $B_s\to \phi\nu\bar\nu$, and
$\Lambda_b \to \Lambda\nu\bar\nu$ decays, and maybe even $B\to \pi(\rho)\nu\bar\nu$.  

The two-body decays $B\to\ell^+\ell^-$ are sensitive to particularly
high scales among $B$ decays, especially if BSM physics alleviates the helicity suppression in the SM.  From an effective theory point of view, these decays have some
of the highest mass-scale sensitivities, comparable to $K\to\pi\nu\bar\nu$.  The \fccee is expected to be comparably sensitive to the HL-LHC for the decays
$B_{s,d}\to \mu^+\mu^-$, but should be a lot more sensitive for $B_{s,d}\to e^+e^-$, and measure $B_s\to \tau^+\tau^-$ even at the SM level, ${\cal B}(B_s\to \tau^+\tau^-) =
(7.7\pm0.5)\times 10^{-7}$~\cite{Bobeth:2013uxa} (with 80
events/exp/year~\cite{DonalHill}).  

In many models inspired by the $R_{K^{(*)}}$ and $R(D^{(*)})$ anomalies hinting at lepton universality violation, there
are correlated deviations from the SM predictions in transitions mediated by
operators with flavor structures $b\bar s \ell^+\ell^-$ and $b\bar s
\nu\bar\nu$, which makes these modes particularly interesting.  
If any of the anomalies become established, then searches for $b\to s\tau\mu$, $b\to s\tau e$, and similar modes would gain a lot in importance, since lepton universality violation in most models also leads to lepton flavor violation.
The not yet observed $B_c\to\tau\bar\nu$ decay, which the \fccee should be able to measure~\cite{Amhis:2021cfy}, is impacted by many models that attempt to explain the $R(D^{(*)})$ anomaly.  
The challenge to precision physics using $B_c$ decays may be the knowledge of production rates.
These decay modes are important even if the current anomalies become less significant,
as they are generic probes of many BSM scenarios~\cite{Cohen:1996vb} which treat the 3rd
generation differently from the 1st and 2nd.

In order to fully benefit from these measurements, precise
SM calculations of these rates are needed. Regarding the inclusive calculations, 
the operator product expansion will probably remain the main tool.
The calculations for exclusive decays will likely rely on lattice QCD, and new developments are needed to extend the calculations for the full ranges of $q^2$. 
Another significant challenge for lattice QCD calculations is to account for QED corrections and isospin violation.  These have been started to be addressed in limited contexts, and they will have to be included more comprehensively~\cite{Sachrajda}.
Fully addressing electromagnetic corrections will necessitate dedicated interactions between theorists and experimentalists, for example to refine Monte Carlo tools. 
The hope is that by the time data from \fccee\ is available, the theoretical uncertainties may be below the anticipated experimental ones.

\subsection{Polarized baryons and quarks}

Many probes of flavor that can be done with mesons can also be done
with baryons. In many cases, adding baryons provide more statistics as well
as a different set of systematics. There are, however, some ways
baryons can probe short-distance physics that cannot be done with mesons.
In particular, baryons can be polarized, which is not the case for the
pseudoscalar mesons that are most often used to probe the weak
interaction. The fact that $b$ and $c$ quarks in $Z$ decays are
highly polarized, make polarization related physics
well suited for the \fccee.

There are two aspects of polarization that can be exploited. First, by
determining the polarization of the baryons, we can learn about 
the polarization of the underlying quark, which can teach us about the
production
mechanism~\cite{Falk:1993rf,Galanti:2015pqa}. In particular, it can
teach us about the Dirac structure of the operator that creates
them. That information is washed out by hadronization into pseuduscalar mesons.
(For example, if squarks are discovered, we would like to know if they are the left- or the right-handed ones. One way to probe this is to measure the polarization of the
quarks that emerge from their decays.) In order to be able to use
baryon polarization as a measure of the quark polarization, we need to know how much of the quark polarizations is retained by the baryons.
It was shown that this can be obtained using top decays~\cite{Galanti:2015pqa}. 
A generalization of that idea to $Z$ decay is needed
in order to fully exploit this
method at the \fccee. Moreover, comparing the two determinations is
another interesting check of the SM.

Polarization is also required in order to explore the full
structure of the weak decays of the quarks.
The point is that
$\Lambda_b$ baryons produced at the \fccee are highly polarized. The reason is that in
$Z \to b \bar b$ the quarks are highly polarized and based
on data from LEP~\cite{Buskulic:1995mf,Abbiendi:1998uz,Abreu:1999gf} 
and theoretical estimates~\cite{Falk:1993rf,Galanti:2015pqa}, we expect an
${\cal O}(1)$ fraction of that polarization to be retained by the $\Lambda_b$. Thus, we can have a very large sample of polarized $\Lambda_b$ decays. 
For example, looking for semileptonic $\Lambda_b\to
\Lambda_c \ell \nu$ decays with polarized $\Lambda_b$ can test the handedness
of the weak interaction in similar ways that it is done with the
Michel parameters in muon decays~\cite{Manohar:1993qn}. It can also be used to
test the structure of FCNC decays, like $\Lambda_b \to \Lambda
\ell^+\ell^-$~\cite{Das:2020qws}. Similar studied can be done for
$\Lambda_c$ decays~\cite{deBoer:2017que}.

\subsection{Exclusive hadronic $Z$ decays}

One way to learn about some nonperturbative aspects of QCD is to study meson
productions in exclusive $Z$ decays. For example, the rate of $Z\to
X\gamma$ is sensitive to the light-cone distribution amplitude of the
meson $X$~\cite{Grossmann:2015lea}. This avenue to 
probe these amplitudes is very promising
theoretically, as the expansion parameter is $1/Q$, where $Q$
is the energy carried by the hadron. Thus, using $Z$ decays, we can get
theoretical sensitivity that is absent in $B$ decays. A promising
candidate for such decay is $Z\to J/\psi\gamma$, where the
branching ratio is expected to be of ${\cal O}(10^{-7})$. 
Thus, at the \fccee, we can hope
to measure the rate with high precision.

The interest in these decays goes beyond QCD as they are closely
related to decays of the form $H\to X\gamma$. Such decays can probe
the couplings of the Higgs to light
quarks~\cite{Perez:2015aoa}. These couplings are very hard to measure,
and exclusive decays may be the best option. 
Thus, $Z$ decays provide important calibration for these calculations.

Another unique opportunity of the \fccee is to look for FCNC $Z$ decays, e.g., $Z\to B_s \gamma$ or $Z\to B_s \mu^+ \mu^-$. 
In the SM such decay rates are expected to be suppressed compared
to $Z\to J/\psi\gamma$ by roughly $[|V_{cb}|/(16 \pi^2)]^2 \sim
10^{-7}$, resulting in branching ratios smaller than about $10^{-14}$,
and thus too small to measure. Yet, one can envision a situation
where such rates are enhanced by some BSM physics. In many cases, such
an enhancement is ruled out by rare $B_s$ and $B$ decays. Yet, there
could be cancellations between several contributions in $B$ decays that
is absent in $Z$ decays. 
While we are not aware of any model that
predicts such an enhancement, the \fccee is
unique in its ability to probe directly such FCNC couplings of the $Z$ boson that
is not available from $B$ decays. Theoretically, it would be
interesting to study decays like $Z \to B^+ K^-$ or $Z \to B^+ K^-
\gamma$ to check if they can provide sensitivity to FCNC $Z$ couplings.

\subsection{Charm physics}
\label{sec:bsm.charm}

$CP$ violation in both charm mixing and decay are probes of QCD dynamics and BSM physics.
In 2011, the LHCb hint of a nearly 1\% $CP$ violation in $D$ decays
generated intense attention, 
since the central value was larger than what was previously considered to be allowed in the
SM~\cite{Isidori:2011qw,Brod:2011re,Pirtskhalava:2011va,Brod:2012ud}. In
2019 a smaller central value was established with more than $5\sigma$
significance, 
$A_{CP}(K^-K^+) - A_{CP}(\pi^-\pi^+) = -(1.54\pm0.29)\times 10^{-3}$~\cite{Aaij:2019kcg}.
It is possible to accommodate this result in the SM, but it requires hadronic enhancements compared to the
available (model dependent) calculations.  It is expected that the \fccee will be able to measure many individual $CP$ asymmetries with good precision, without having to construct differences, like the one recently measured by LHCb.

The \FCCee can probe many more observables that can teach us about CPV
in the charm system. In particular, we should be able to have many tests of the
SM, as we briefly explain below. These tests can either confirm the
picture that there are large hadronic enhancements in certain channels, or will show
that there is some BSM physics that is significant in the $D$ system.

The \FCCee can provide many such measurements. In particular, the
anticipated superb time resolution can make them more precise than we can hope
to get at the LHC and Belle~II. Moreover, one problem for the LHC
when it attempts to probe very small $CP$ asymmetries has to do with the 
production
asymmetry between $c$ and $\bar c$, which can only be controlled using other measurements. This issue is not there in the
\FCCee, as the production is symmetric.

While we cannot make precision calculations for charm CPV because of 
hadronic uncertainties, we are still able to make rough predictions
that can be tested. In particular, one can explore consequences of flavor $SU(3)$ symmetry. The point is that different observables arise
at different orders of $SU(3)$ breaking, and while we cannot calculate how
large this breaking is in specific matrix elements, it is typically of order $20\%$. Thus, by
exploiting relations that hold to higher order in $SU(3)$ breaking, we can
derive a pattern that can be used to probe the SM~\cite{Kagan:2020vri,Grossmann:2021aaa}.

In particular, the SM predicts that to leading order in $SU(3)$
breaking all the time-dependent $CP$ asymmetries are
identical. Moreover, there are subsets of them that are 
identical to second order in the breaking. To test
these predictions, we need to have many different measurements, with
uncertainties that are at the level of the anticipated $SU(3)$ symmetry breaking
effects.

The situation can dramatically change if there is a theoretical
breakthrough that enables the calculation of some of the hadronic
input needed for these observables. The prime candidate is
lattice QCD, and while some preliminary studies for the matrix
elements relevant for $A_{CP}(K^-K^+) - A_{CP}(\pi^-\pi^+)$ have started,
it is yet unknown what can be achieved for these observables.

\section{Conclusions}

Flavor physics is a mature field, with a lot of promising directions
to probe BSM physics and QCD dynamics.  It has played critical roles in developing the SM, and provides some of the strongest constraints on BSM physics. The data from the \fccee will be important in moving this program forward. The field of particle physics when the \fccee starts may be rather different from where we
are today, but it is clear that there are many ways \fccee can shed
light on open questions in flavor physics after the end of the LHC and Belle~II. In this short paper we tried to touch upon some less frequently discussed
areas, to show the breadth of topics where the \fccee can make a significant impact.

Some advances are needed in theoretical particle physics in order to take full
advantage of the \fccee data. Improvements are expected in lattice
QCD, which will enable better control over hadronic
uncertainties in certain processes.  We also anticipate new calculations of
higher-order corrections to precision measurements. 
Theoretical advances are also needed in understanding hadronic physics in some areas where we do not yet know how to make progress using
current effective field theory or lattice QCD methods.  
While we cannot guess how breakthroughs will arise and what their
scope may be (in terms of matrix elements that the novel methods can better calculate), large new data sets in the past have always resulted in unexpected developments. 
Whatever the situation will be when data from the \fccee is analyzed, we are going to learn significant 
new information about flavor physics from them.

\subsection*{Acknowledgements}

We thank Stephane Monteil, Dean Robinson, and Marie-Helene Schune for helpful
discussions.
The work of YG is supported in part by the NSF grant PHY1316222.
ZL was supported in part by the Office of High Energy Physics of the 
U.S.\ Department of Energy under contract DE-AC02-05CH11231.

\bibliography{fccflavor}

\end{document}